\documentclass[prb,preprint]{revtex4}
\usepackage{epsfig}

\begin{document}

\draft

\title{Structure and energetics of Ni clusters with up to 150 atoms}

\author{Valeri G.\ Grigoryan\footnote{Corresponding author. e-mail:
vg.grigoryan@mx.uni-saarland.de; fax: +49 681 302 3857} and Michael Springborg\footnote{e-mail:
m.springborg@mx.uni-saarland.de; fax: +49 681 302 3857}}
\affiliation{Physical Chemistry, University of Saarland,
66123 Saarbr\"ucken, Germany}

\date{\today}

\begin{abstract}

We present a method (the {\sl Aufbau/Abbau} method) 
for optimizing the structure of a whole series of clusters
without making any assumptions on the structure. Subsequently, the method is combined
with the embedded-atom method in determining the structure of the two energetically
lowest isomers of Ni$_N$ clusters with $N$ up to 150. Finally, various analytical
descriptors are introduced that are used in studying the overall shape of
the clusters, their structure and stability, and possible growth and dissociation processes.

\end{abstract}

\maketitle

About two decades ago the experimental observation and theoretical description of 
so-called magic numbers in clusters of simple metals (see [\onlinecite{deheer,brack}])
provided an excellent example of how experiment and theory could supplement each other
in the development of the understanding of materials properties. The theoretical studies 
for any but the smallest clusters were based on the jellium model, i.e., the internal
structure of the cluster was completely neglected and the core electrons and the nuclei
were smeared out to a homogeneous jellium in whose field the valence electrons moved.
Moreover, in the first studies the jellium was supposed to be spherical and to have sharp
boundaries, although these assumptions in later studies have been relaxed 
partially (see [\onlinecite{brack}]).

For theoretical studies of any other metal clusters M$_N$ than those of simple metals,
one encounters the serious problem that when the valence electrons are anything but 
completely delocalized, the (spherical) jellium model may not provide a realistic 
approximation. This is, e.g., the case for transition-metal clusters where the nearest-neighbor
interactions between the atoms are important, so that the precise structure of the cluster
as well as the inhomogeneities of the electron density are to be included in any realistic
description. Then, furthermore, the fact that the number of structural degrees of freedom ($3N-6$ or
$3N-5$) together with the fact that the number of metastable structures grows very
fast with $N$ makes detailed systematic studies of how the properties 
depend on $N$ for any but the absolutely smallest clusters extremely demanding. On the
other hand, the further development of this
field requires a precise understanding of how the properties of interest depend on
the structure, size, and composition of the clusters, which, unfortunately, is
very difficult to achieve theoretically or experimentally.

Here we shall report results of a theoretical study of the structural and
energetic properties of Ni$_N$ clusters for all $N$ up to 150. Our calculations have made 
no assumptions about the structure of the clusters whereby they differ from
most other studies of these systems (see, e.g., [\onlinecite{doye,luo,mc,boy,wetzel}] 
and references
therein) of which the previously most detailed and unbiased study\cite{doye} considered only up
to $N=80$. Since we have had to perform
often several 1000 structure-relaxations for a given $N$ without making any assumption
on structure or symmetry, parameter-free methods cannot be applied, and, therefore,
we have employed the embedded-atom method (EAM) 
in the calculation of the total energy for a given structure. Moreover, we have
developed an {\sl Aufbau/Abbau} method for the structure optimization. This method is 
partly inspired by experimental conditions where clusters often are produced by adding
atoms individually to a seed, although the experiments often take place at 
conditions far from equilibrium and from $T=0$ K. A preliminary description of
our approach has been presented elsewhere,\cite{pccp,psik} but compared with that 
work the present one is much more
accurate (through a much more detailed structure optimization that also includes
more isomers), and, in addition, we shall here present various theoretical tools with
which we can extract the fundamental structural and energetic properties of 
these clusters.

In our {\sl Aufbau/Abbau}
method we consider clusters with $L$ and $L+P$ atoms with $P\simeq5-10$. For each of those
we study a set of randomly generated structures. Using a quasi-Newton method relaxed
local-total-energy minima are identified and the structures of the lowest total energy selected.
One by one atoms are added to the structure with $L$ atoms (many hundred times for each
size), and the structures are relaxed. Parallel, one by one atoms are removed from the structure
with $L+P$ atoms. From this two series of structures for $L\le N\le L+P$ 
those structures of the lowest energies are chosen
and these are used as seeds for a new set of calculations. First, when no lower total
energies are found, it is assumed that the structures of the global-total-energy minima have
been identified, and we proceed to larger clusters. In some few cases, extra 
high-symmetry structures were added as possible structures, which in one case
($N=75$) resulted in an otherwise unidentified lowest-energy structure. According to our 
experience, for a given $N$ we have to consider about 500 randomly
generated structures for the smallest values of $N$ and about 1000 for the
largest values. Moreover, about 3 {\sl Aufbau/Abbau} cycles are needed for
the smaller $N$ and up to 10 for the larger $N$. 

In comparison with our
earlier study, the method has been extended so that by keeping track of more 
structures of the lowest total energies, we obtain the energetically
lowest isomers for each size, and a much more detailed search has led to the 
identification of new structurals for several $N$. Since our approach (as any other unbiased method for structure
optimization of larger, low-symmetric systems) requires a large number of total-energy
calculations for different, larger structures, it cannot be combined with accurate 
parameter-free electronic-structure methods, but has to be based on approximate schemes
for calculating the total energy. Here, we combine the {\sl Aufbau/Abbau} method
with the EAM method for the latter purpose.

The basic idea of the EAM\cite{eam1,eam2,eam3,eam4} is to consider
every atom as an impurity embedded in a host provided by all other atoms. The tails of 
the electron densities of the host atoms inside the guest atom lead to a modification
in the energy of this atom, $F_i(\rho_i^h)$, where $i$ labels the atom, and $\rho_i^h$
is the electron density from the host atoms. Usually this is evaluated at the site
of the nucleus of the $i$th atom. The remaining part of the binding energy of the system 
of interest is
approximated through pair potentials, $\frac{1}{2}\sum_{j\ne i}\phi_{ij}(R_{ij})$ with
$R_{ij}$ being the distance between atoms $i$ and $j$. The embedding energies $F_i$ and
the pair potentials $\phi_{ij}$ have been obtained by fitting to theoretical and experimental
information on different infinite systems.\cite{eam3,eam4} The 
EAM is conceptually simple, but it grasps the major parts of
the interatomic interactions and gives accurate structural and energetic
information on various extended and truncated systems.\cite{eam3,eam4,mishin} 

\begin{figure}[ht]
\begin{picture}(18,12)
\put(0,05){\psfig{file=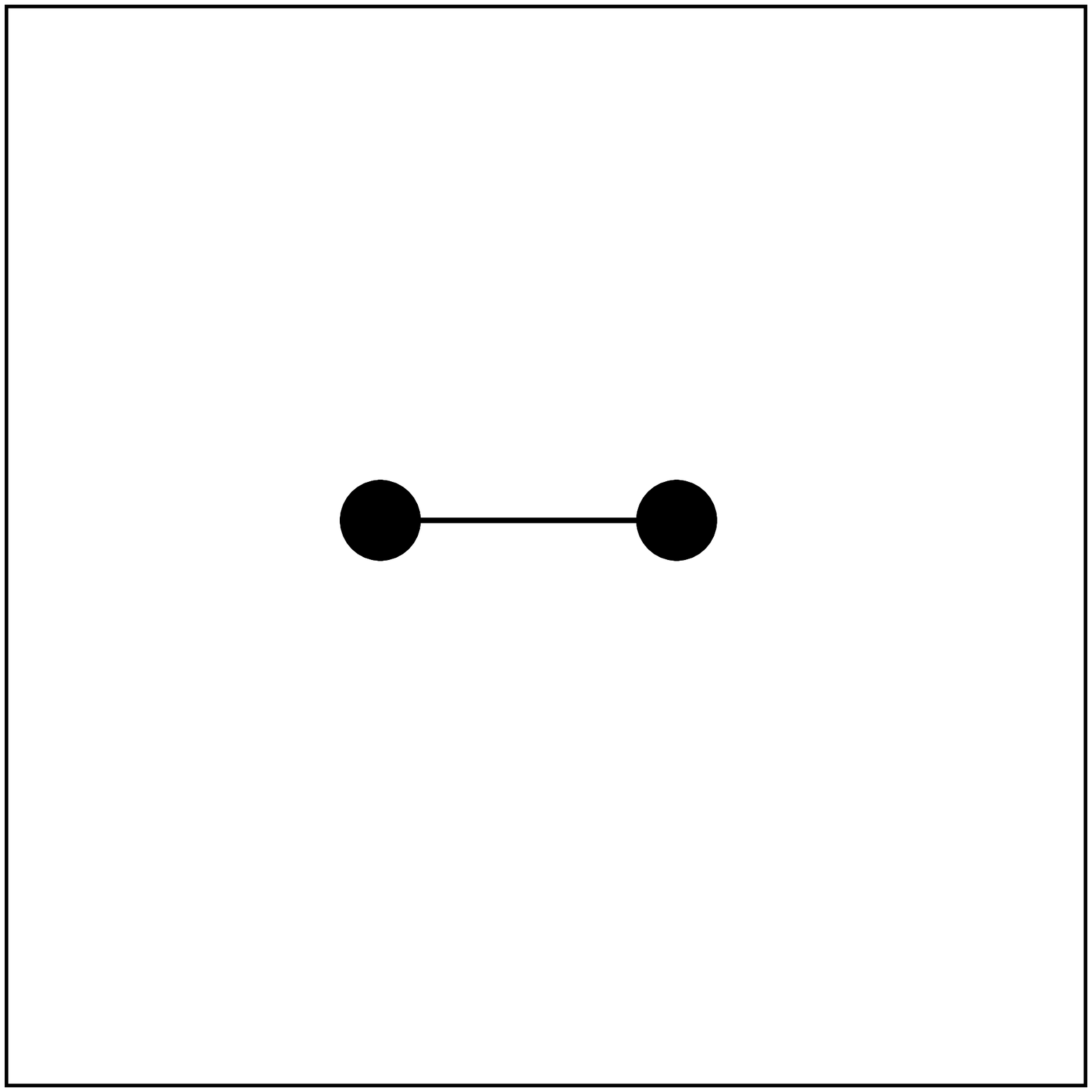,width=3cm}}
\put(1.2,04.5){Ni$_{2}$}
\put(4,05){\psfig{file=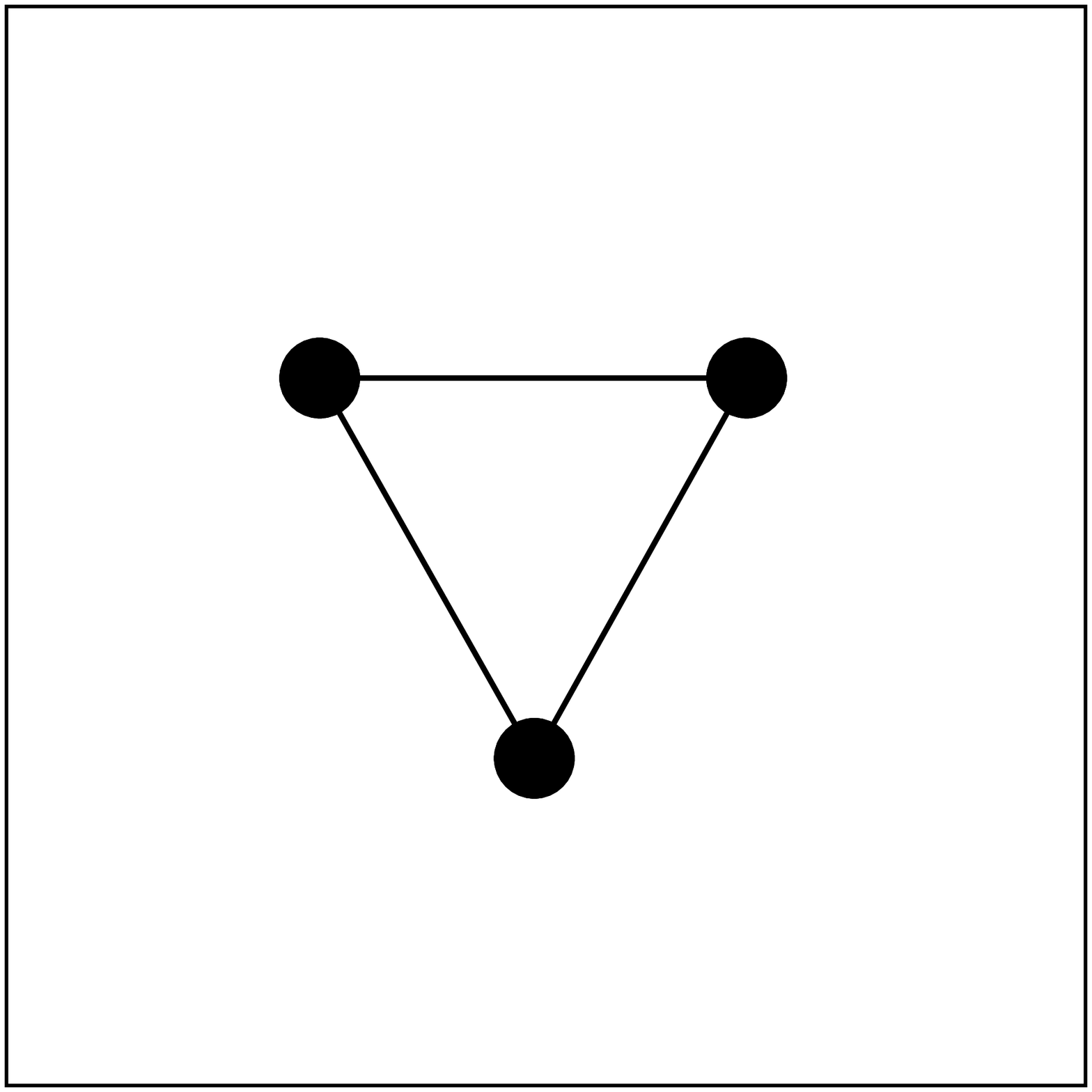,width=3cm}}
\put(5.2,04.5){Ni$_{3}$}
\put(8,05){\psfig{file=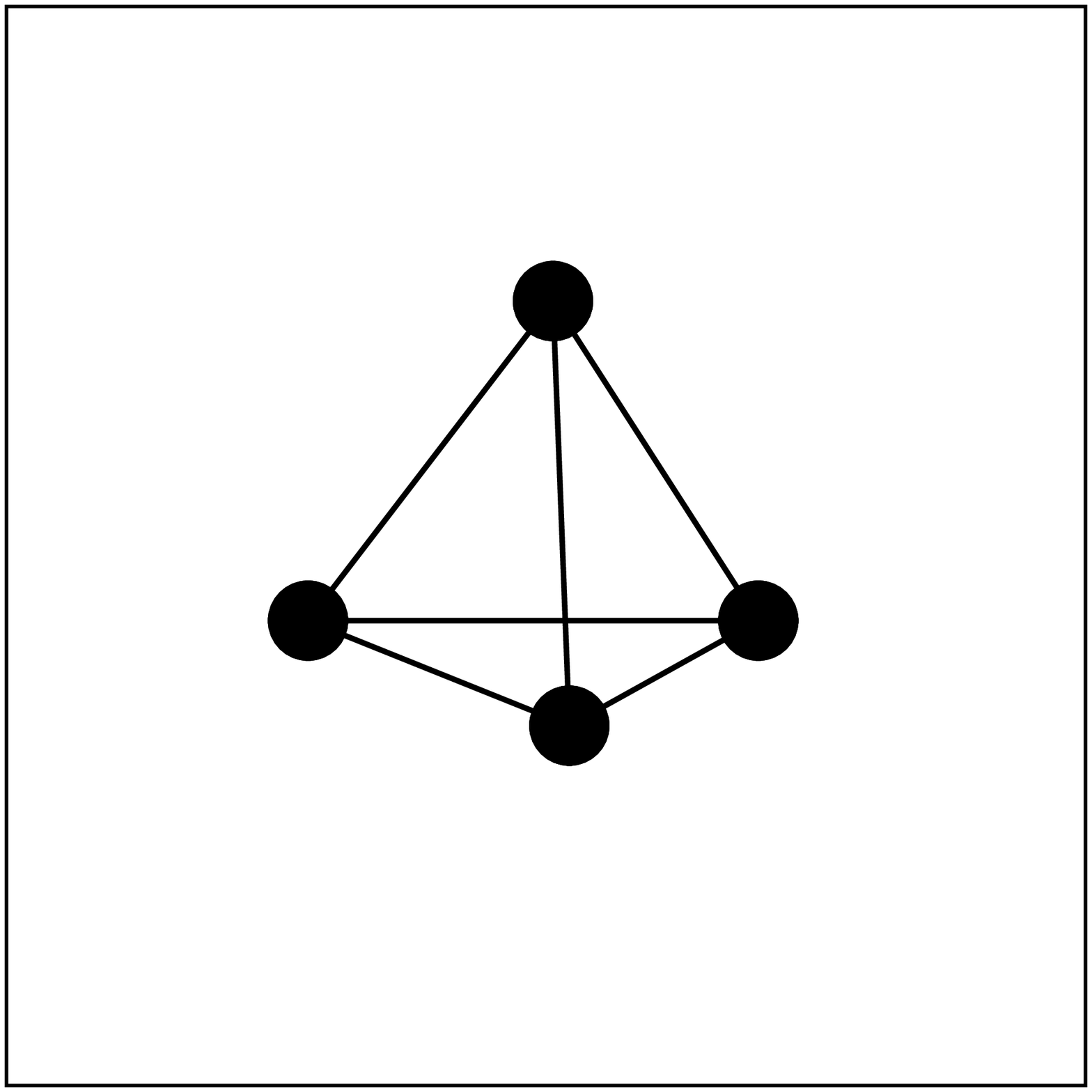,width=3cm}}
\put(9.2,04.5){Ni$_{4}$}
\put(12,05){\psfig{file=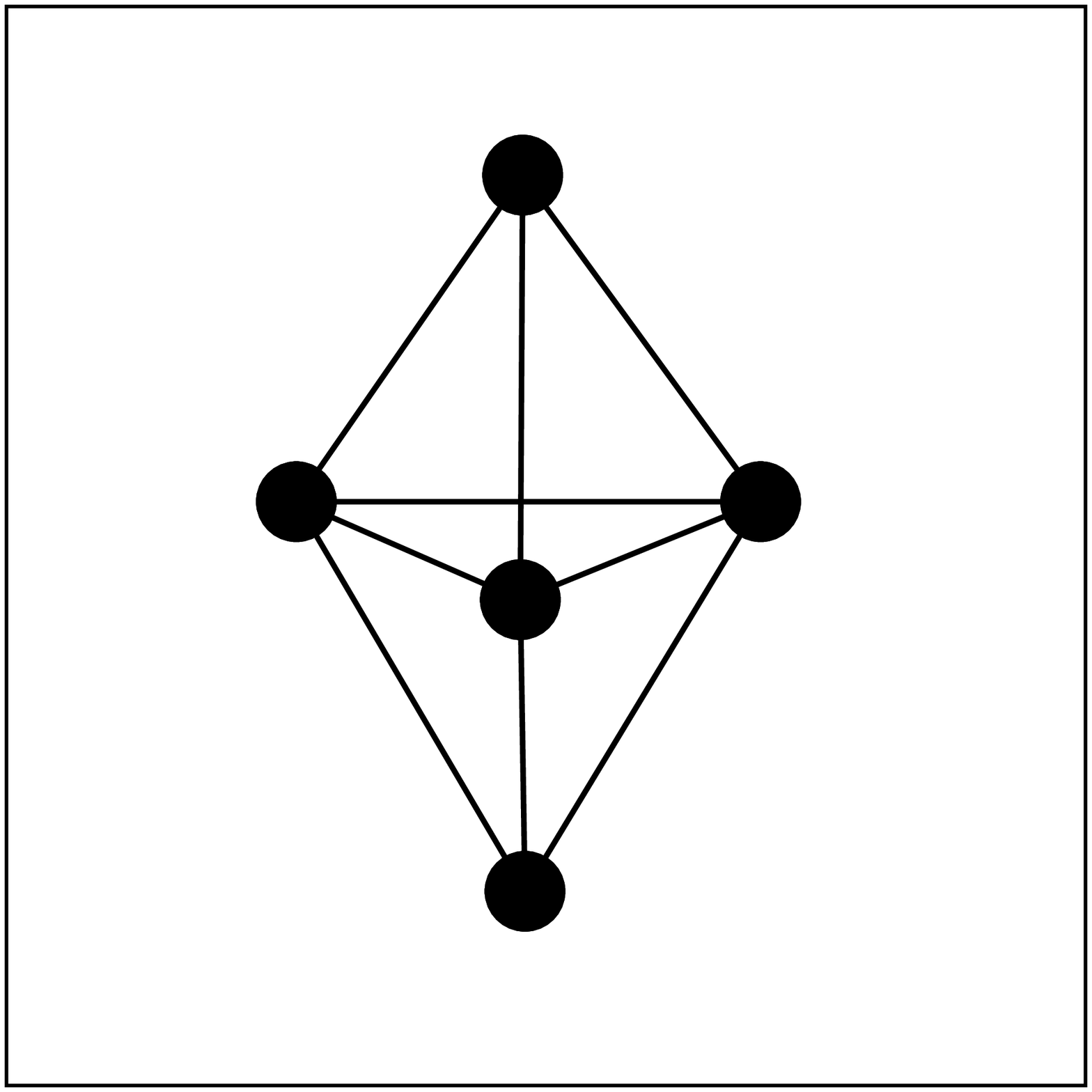,width=3cm}}
\put(13.2,04.5){Ni$_{5}$}
\put(0,01){\psfig{file=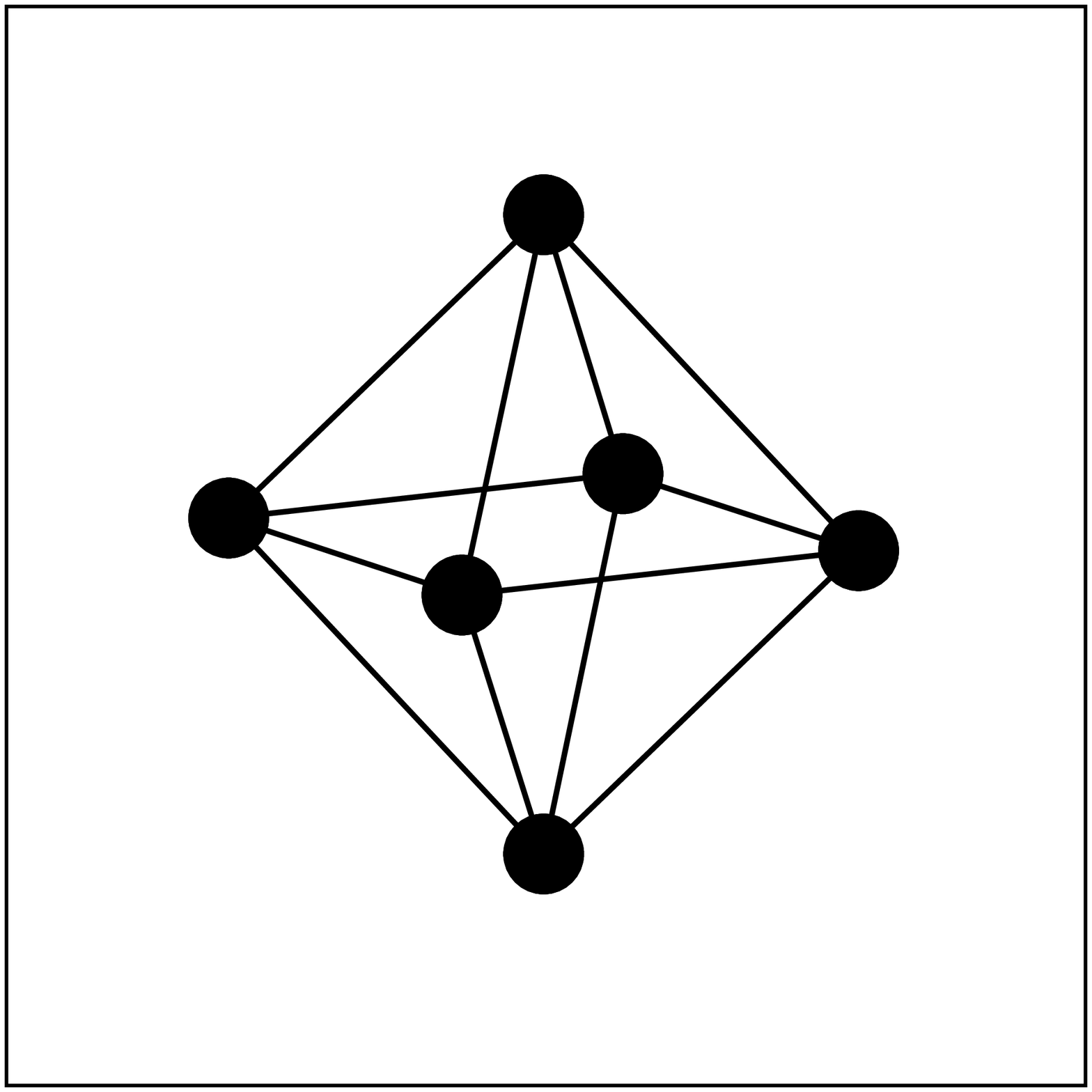,width=3cm}}
\put(1.2,00.5){Ni$_{6}$}
\put(4,01){\psfig{file=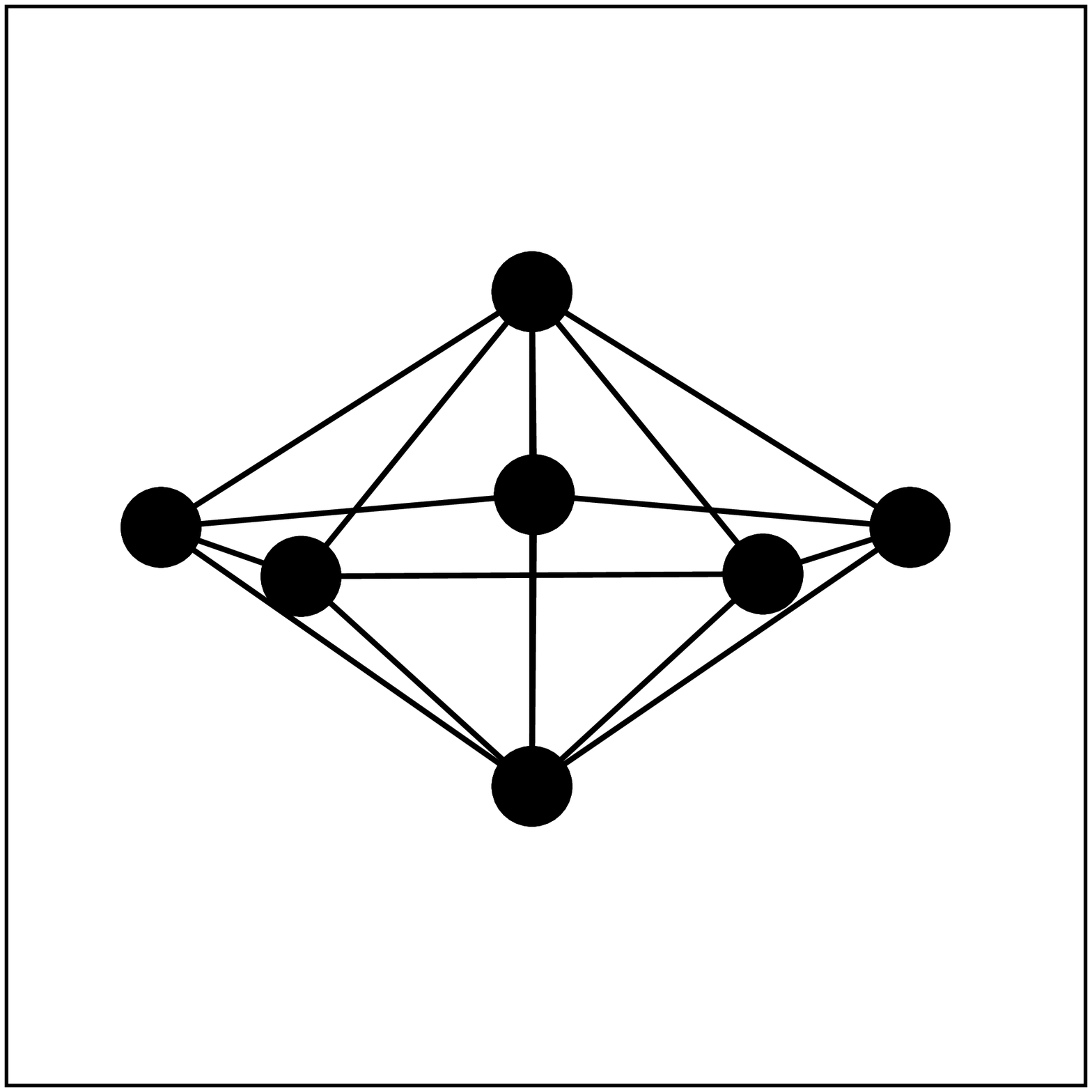,width=3cm}}
\put(5.2,00.5){Ni$_{7}$}
\put(8,01){\psfig{file=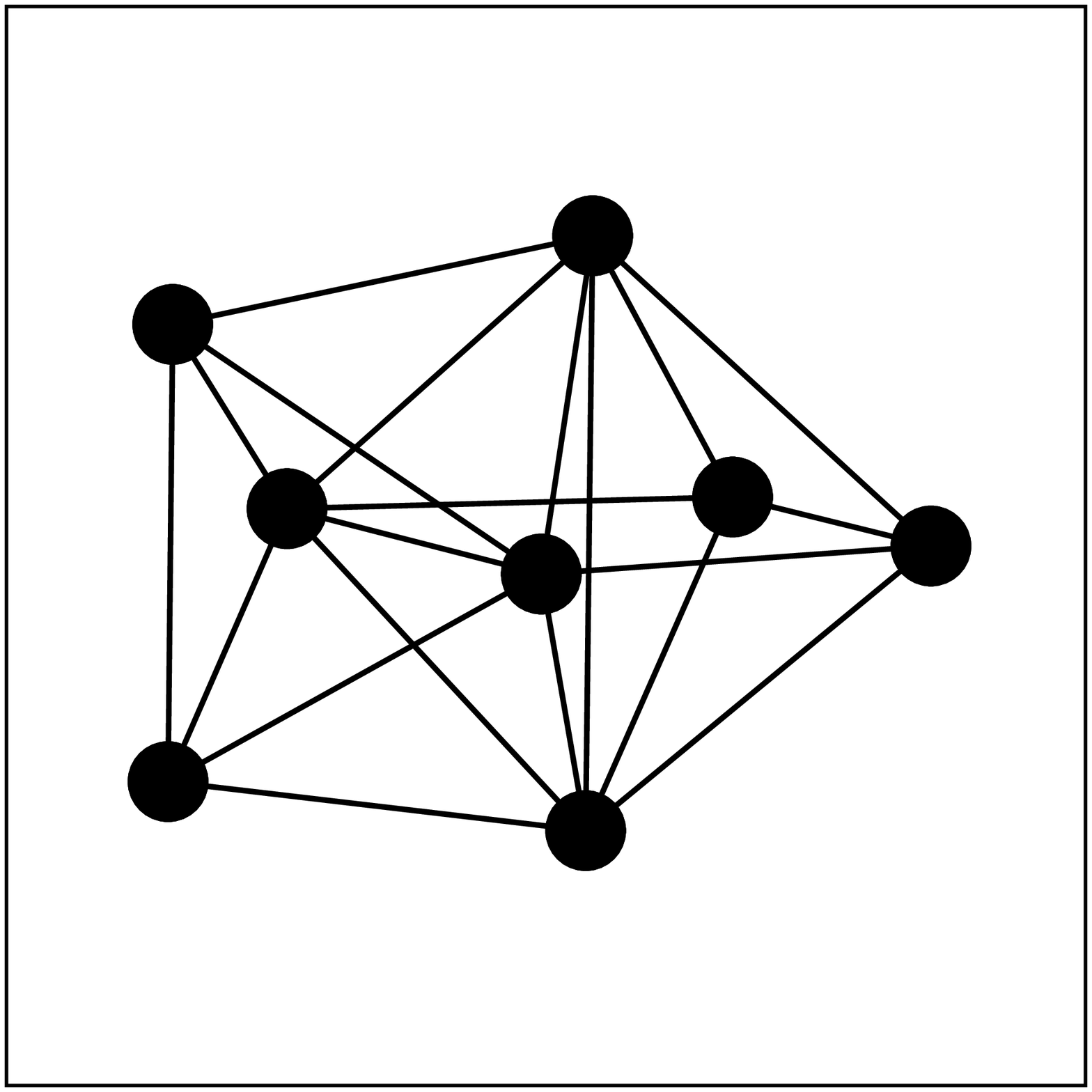,width=3cm}}
\put(9.2,00.5){Ni$_{8}$}
\put(12,01){\psfig{file=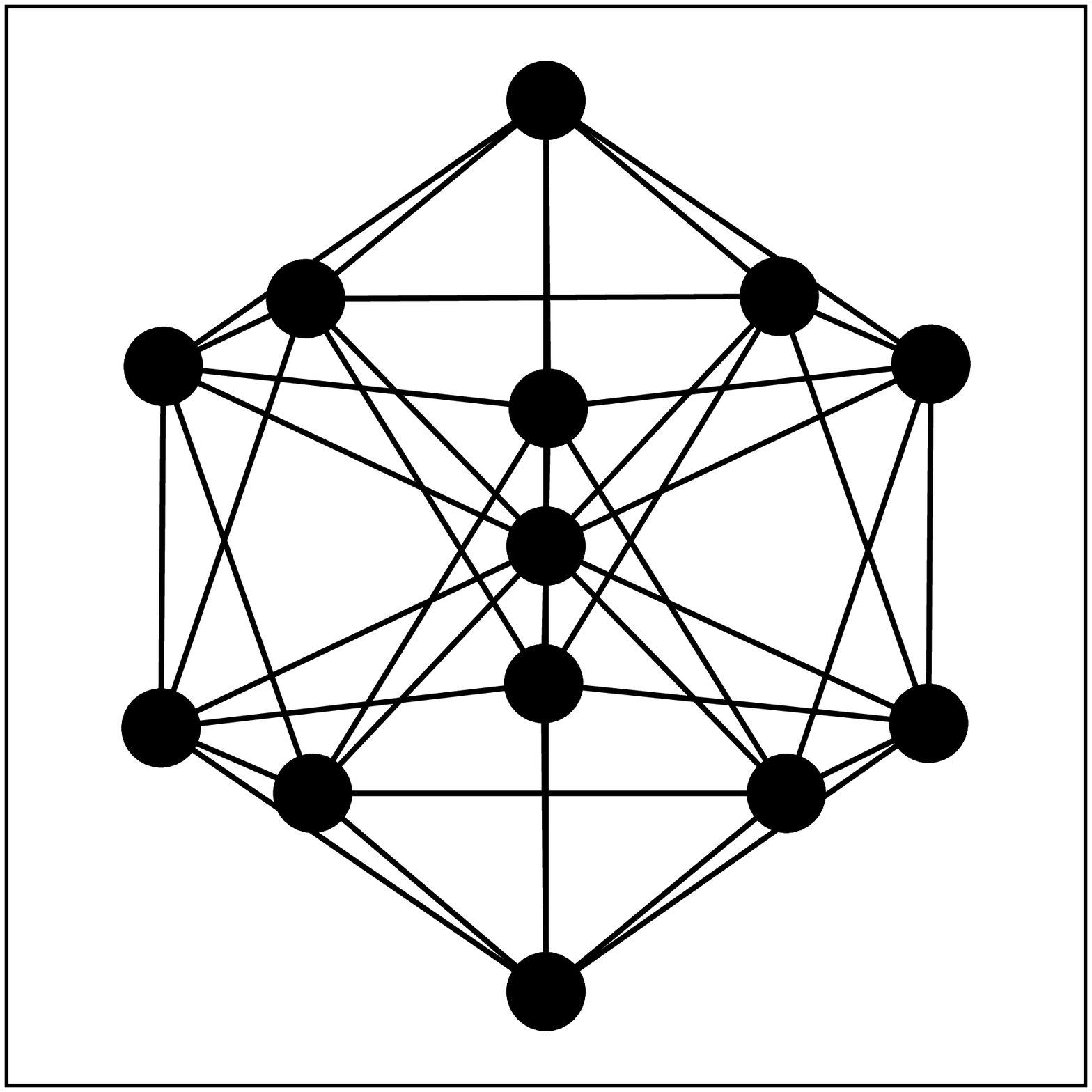,width=3cm}}
\put(13.2,00.5){Ni$_{13}$}
\end{picture}
\caption{The optimized structure of the nickel clusters listed in Table I.}
\label{fig:fig00}
\end{figure}

Before discussing our results in detail we shall address the question of the accuracy of
the EAM method. As such, it is constructed first of all for describing cohesion in
extended metallic systems like crystals without or with defects and surfaces. Due to
the complexity of the present study, it is not possible to apply a parameter-free
method, and simpler methods like the EAM have to be used, but nevertheless it may
be asked whether the EAM can be used for the, after all, relatively small systems of our
present study. We shall therefore compare our calculated structural properties for
nickel clusters with $N=$ 2, 3, 4, 5, 6, 7, 8, and 13 with those of accurate electronic-structure
calculations using different density-functional methods.\cite{calleja,michelini,desmaris,krueger,reuse}
Table \ref{comparison} list some of the key quantities in the structural characterization
of the different clusters as obtained with the different methods, and in
Fig.\ \ref{fig:fig00} our optimized structures are shown. It is seen in
Table \ref{comparison} that our
results agree well with those of the other studies, both concerning the structure itself
and concerning the interatomic distances. Most of the discrepancies are related to
smaller distortions that may be due to electronic effects (e.g., Jahn-Teller 
distortions) that are not included in the EAM method. Thus, we conclude that our
approach is accurate.

\begin{figure}[ht]
\begin{picture}(15,20)
\put(0,1){\psfig{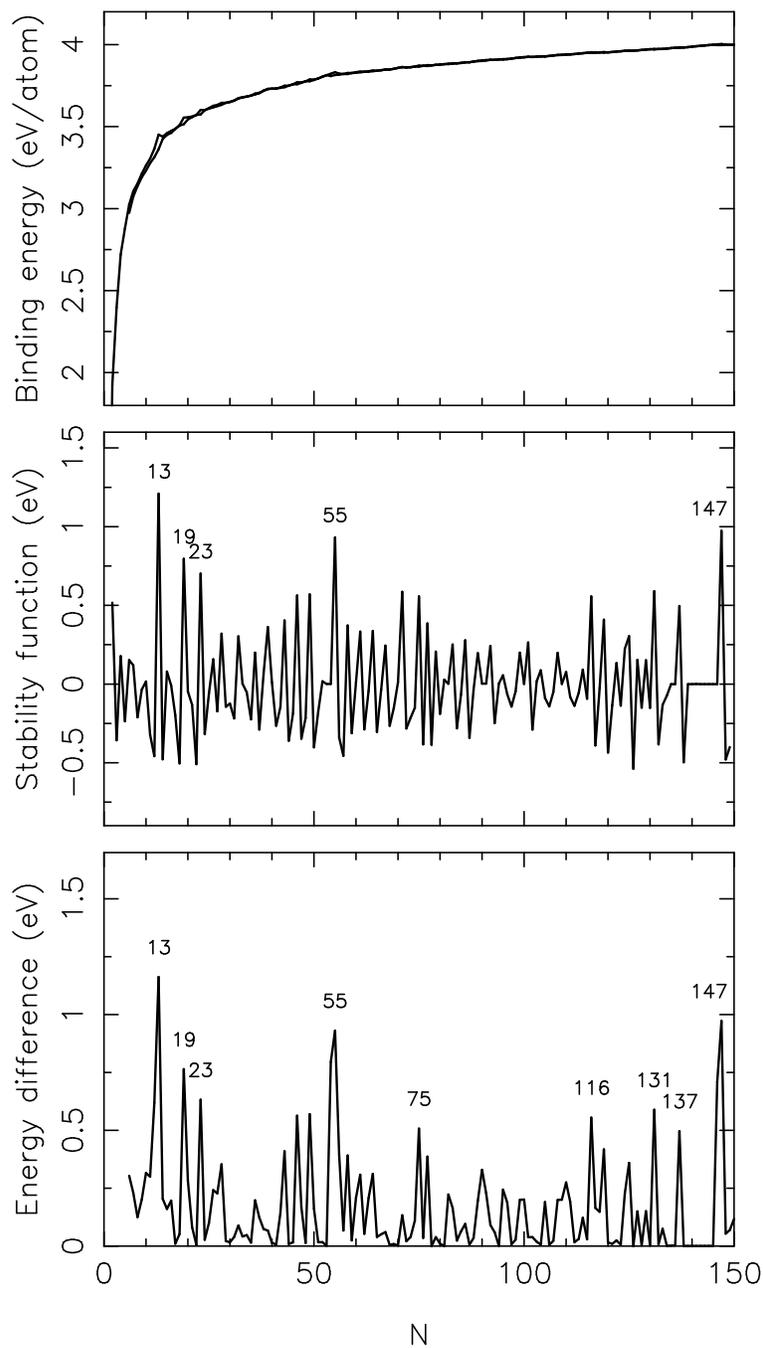}}
\end{picture}
\caption{The binding energy per atom for the energetically two lowest
isomers (upper part), the stability function
(middle part), and the total-energy difference for the two energetically
lowest isomers (lowest part) for Ni$_N$ clusters as functions of $N$.}
\label{fig:fig01}
\end{figure}

In Fig.\ \ref{fig:fig01} we show our calculated binding energy per atom for nickel clusters
with $N$ ranging from 1 to 150. As is generally found, the binding energy for clusters is first
of all a monotonously increasing function of $N$ with, however, some extra structure. In order
to identify the latter more clearly, we show in the figure also the stability function $S(N)=
E_{\rm tot}(N+1)+E_{\rm tot}(N-1)-2E_{\rm tot}(N)$ which has maxima for particularly
stable structures. These are also marked in the figure. As we shall see, some of the 
particularly stable structures correspond to particularly symmetric ones. Particularly
stable structures are found for $N$ = 13 (the most pronounced one), 19, 23, 39, 55, and
147. 

There exists some previous experimental and theoretical studies of the properties
of Ni$_N$ clusters, of which many focus on the inner structural aspects (including
symmetry, see, e.g., [\onlinecite{stave,boy,wetzel,doye,parks1,parks2}]). 
The most detailed study of the energetics of Ni$_N$ clusters
is the theoretical one of Doye and Wales\cite{doye} who used the Sutton-Chen potential in
calculating structural and energetic properties of Ni$_N$ clusters with $N$ up to 80. They
find particular stability for $N=$ 13, 38, 50, 55, 64, 71, 75, and 79, which is in partial
agreement with our results. As discussed by Parks $et$ $al.$\cite{parks1,parks2} the
structures given by Doye and Wales seem to deviate somewhat from experimental findings,
suggesting that the Sutton-Chen potential is not sufficiently accurate for such studies. 
In a tight-binding molecular-dynamics study, Luo studied clusters with up to 55 atoms.\cite{luo}
He found most stable structures for $N=$ 21, 31, 35, 38, 40, 47, 50, and 55, only
partly in agreement with the results of Doye and Wales and of us. On the other hand,
Montejano-Carrizales $et$ $al.$\cite{mc} used the EAM in analysing the so-called umbrella growth
models for icosahedral Ni clusters. Although they thereby make certain assumptions on the
structure their $N$ values for particularly stable structures are in general in
good agreement with ours and with those of
Doye and Wales. This leads support to the umbrella growth model.
This conclusion is further supported below.

The special stability of the cluster sizes mentioned above is supported by comparing
the energies of the two energetically lowest isomers. First for $N=6$ and onwards we 
identify two different isomers, and their total-energy difference is shown in the
lowest panel in Fig.\ \ref{fig:fig01}. Here, the most significant features correlate closely
with those of the stability function.

Besides the stability a number of other issues have attracted significant interest in
cluster physics. These include the structure of the individual clusters, whether growth modes
can be identified, and how the clusters will dissociate. We shall now address these issues.

\begin{figure}[ht]
\begin{picture}(15,15)
\put(0,1){\psfig{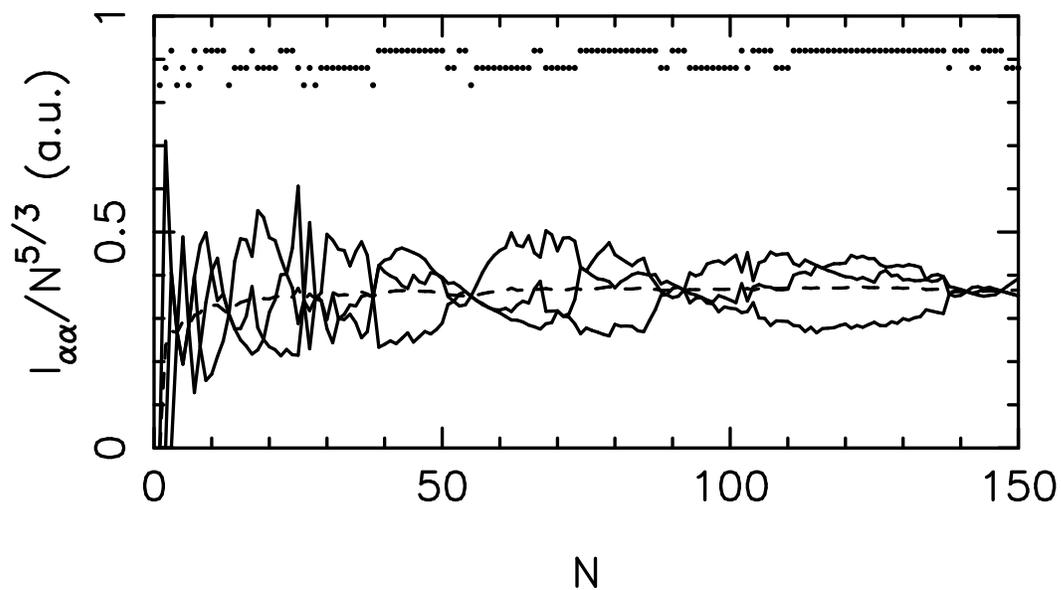}}
\end{picture}
\caption{The normalized eigenvalues of the matrix containing the
moments of inertia as functions of $N$. The three solid
curves gives the eigenvalues, and the dashed curve their average. In the upper part
of the figure a simple estimate of the overall shape of the clusters is given:
either being spherical (marked with points in the lowest row), cigar-shaped (points
in the middle row), or lens-shaped (points in the uppermost row).}
\label{fig:fig02}
\end{figure}

For a given cluster Ni$_N$ we calculate the eigenvalues $I_{\alpha\alpha}$ of the 
matrix containing the moments of inertia $\sum_{i=1}^N s_i t_i$ with $s$ and $t$ 
being the $x$, $y$, and $z$ in a coordinate system with the origin at the center of
mass of the cluster. For a spherical jellium $I_{\alpha\alpha}\propto N^{5/3}$, and therefore
we show (Fig.\ \ref{fig:fig02}) $I_{\alpha\alpha}/N^{5/3}$. If the three eigenvalues
are identical the overall shape of the cluster is approximately
spherical. Such clusters are marked with points in the lowest row in the
uppermost part of the figure and are found for $N$ = 1, 4, 6, 13, 26, 28, 38, and 55, which 
partially correspond to particularly stable clusters; 
cf.\ Fig.\ \ref{fig:fig01}. When two of the eigenvalues are small and only one
is large, the cluster is essentially prolate (cigar-shaped), whereas it with two large
eigenvalues and one small one is oblate (lens-shaped). In the figure we have
marked which of the clusters have these two types of overall shapes. Except for the 
absolutely smallest clusters, we have that,
if the cluster structures can be considered as obtained
by adding atom by atom to a largely fixed core, the clusters will either be close to
spherical (so that several changes between prolate and oblate structures are possible),
or one or the other of the two structure types will dominate in large $N$-intervals. 
However, the results of Fig.\ \ref{fig:fig02} show that except for $N$ 
above about 137, the eigenvalues $I_{\alpha\alpha}$ 
are very different, so that the clusters
have pronounced prolate or oblate shapes. Moreover, several changes between these
two structure types occur in the range of $N$ we have considered. 

A more detailed way of analysing whether the cluster with $N$ atoms can be considered as
formed by adding a single atom to the cluster with $N-1$ atoms is provided by defining
similarity indices as follows. We calculate the $(N-1)(N-2)/2$ interatomic distances $d_{ij}^0$ 
(in a.u.) for the
$(N-1)$-atom cluster and sort them. Subsequently, we consider $all$ the possible $(N-1)$-atomic parts
of the $N$-atom cluster, and for those we also calculate and sort the interatomic distances 
$d_{ij}$. The smallest value of $q_1$ with ${(N-1)(N-2)\over2}q_1^2
=\sum_{i>j=1}^{N-1}(d_{ij}^0-d_{ij})^2$ defines then the
similarity index, $s_1=(1+q_1)^{-1}$, which is 1 for a cluster that is
obtained by adding one atom to the previous one, and approaches 0 for very different
structures. Another similarity index $s_2$ is obtained by instead of using the interatomic
distances we use the distances of the individual atoms to the center of mass of the cluster. Denoting these
so-called radial distances 
$r_i^0$ (in a.u.) for the $(N-1)$-atom cluster and $r_i$ (in a.u.) for the $(N-1)$-atom fragment of the 
$N$-atom cluster, we consider the smallest value of $q_2$ with $(N-1)q_2^2=\sum_{i=1}^{N-1}(r_i^0-
r_i)^2$, and define $s_2=(1+q_2)^{-1}$.

\begin{figure}[ht]
\begin{picture}(10,20)
\put(0,1){\psfig{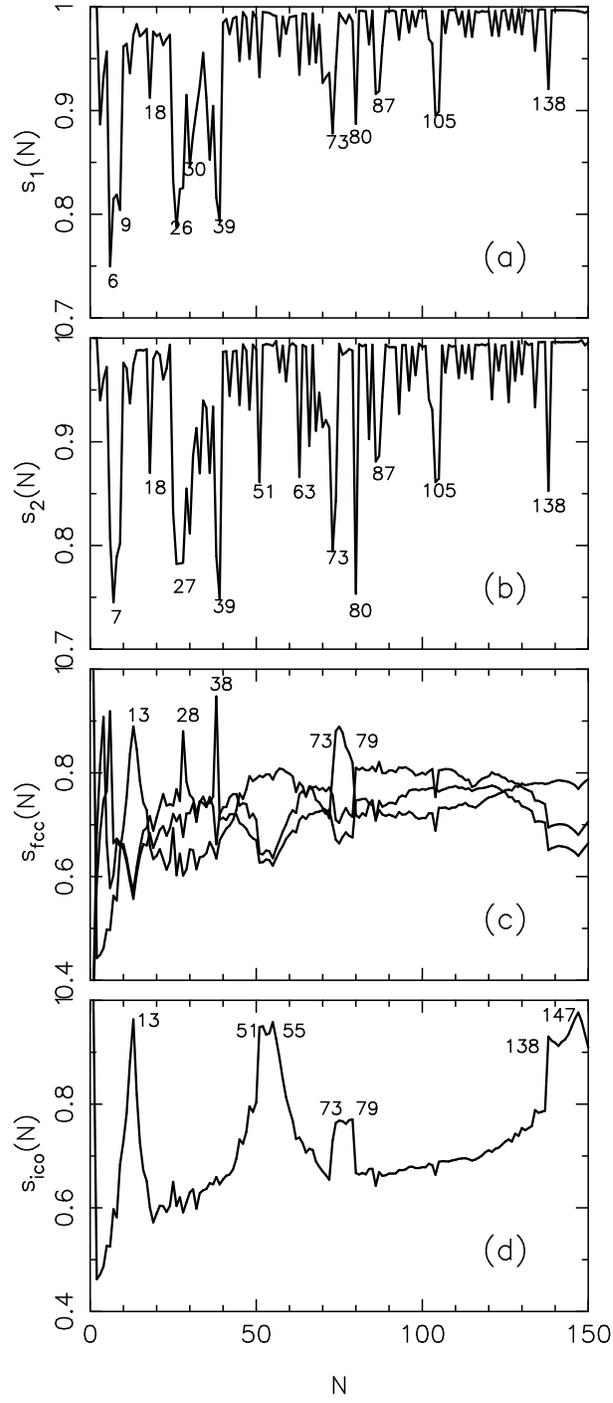}}
\end{picture}
\caption{The similarity indices (a) $s_1(N)$, (b) $s_2(N)$,
(c) $s_{\rm fcc}(N)$, and (d) $s_{\rm ico}(N)$ 
for the Ni$_N$ clusters as a function of $N$. For details, see the text.}
\label{fig:fig35}
\end{figure}

These two functions are shown in Figs.\ \ref{fig:fig35}(a) and (b). Up to $N$ around 40 both
functions possess a number of peaks indicating that no simple growth mechanism can be identified.
This is in accord with the experimental observations of Pellarin $et$ $al.$\cite{pellarin} who
moreover suggest that above that the umbrella growth model agrees with their
observations, in agreement with the results of Montejano-Carrizales $et$ $al.$\cite{mc} 
However, both of our similarity indices find for  larger $N$ peaks at $N=$ 39, 51, 63, 73, 80, 87, 105, 
and 138, implying that also here size-dependent structural differences occur. Thus, in
total we find an overall support for the umbrella growth model although for many
values of $N$ this model does not give an exact description of the structure.

In order to
obtain a quantitative estimate of the similarity of the structure for a given $N$
with that of a fragment of the crystal we proceed as follows. We consider a large, but finite
fragment of the fcc crystal structure and place a center (arbitrarily) at
(0,0,0), (${1\over2}a$,0,0), and (${1\over4}a$,${1\over4}a$,0) (with $a$ being
the length of the unit cell). The set of distances from the atoms to any of these
centers, $\{r_i^{\rm fcc}\}$, is sorted and compared with the radial distances for a
given $N$-atom cluster, $Nq_{\rm fcc}^2=\sum_{i=1}^N(r_i^{\rm fcc}-r_i)^2$. This
defines the fcc similarity index, $s_{\rm fcc}=(1+q_{\rm fcc})^{-1}$. A related 
icosahedral similarity index $s_{\rm ico}$ is defined by using the radial distances
for the (relaxed) $N=309$ icosahedral cluster in comparing with the different
clusters. The two functions are shown in Figs.\ \ref{fig:fig35}(c) and (d). It is 
seen that in the range 73--79 the structure is related both
to that of a fragment of the fcc crystal with, in addition, the center at (0,0,0) and to
the icosahedral structure. This indicates that in exactly this size range more different
structural motifs are present. Two
further peaks can be identified in Fig.\ \ref{fig:fig35}(c): 
at $N=28$ which corresponds to the center at 
(${1\over2}a$,0,0) and at $N=38$ which corresponds to the center at (${1\over4}a$,${1\over4}a$,0). 
The lower panel shows that the icosahedral clusters are important around $N=13$, 
between $N=51$ and $N=55$, and above $N=138$. We finally notice that the structural 
characteristics often are not isolated to singular values of $N$ but can be identified
also around the peaks. 

In order to get insight into possible fragmentation channels we analyse the calculated 
energies $E_{\rm tot}$ as a function of $N$. Neglecting kinetic effects (this is a crude
approximation and may have severe consequences for our conclusions), we seek for a given
$N$ that value of $K$ for which the dissociation energy $E_{\rm tot}(N)-[
E_{\rm tot}(K)+E_{\rm tot}(N-K)]$ is lowest. It turned out that, without 
exception, this value of $K$ was always 0, indicating that any cluster is stabler than its
dissociation products. This result is a consequence of the fact that $E_{\rm tot}(N)$ is
upward curved, as indirectly can be seen in the upper part of Fig.\ \ref{fig:fig01}.

\begin{figure}[ht]
\begin{picture}(15,20)
\put(0,1){\psfig{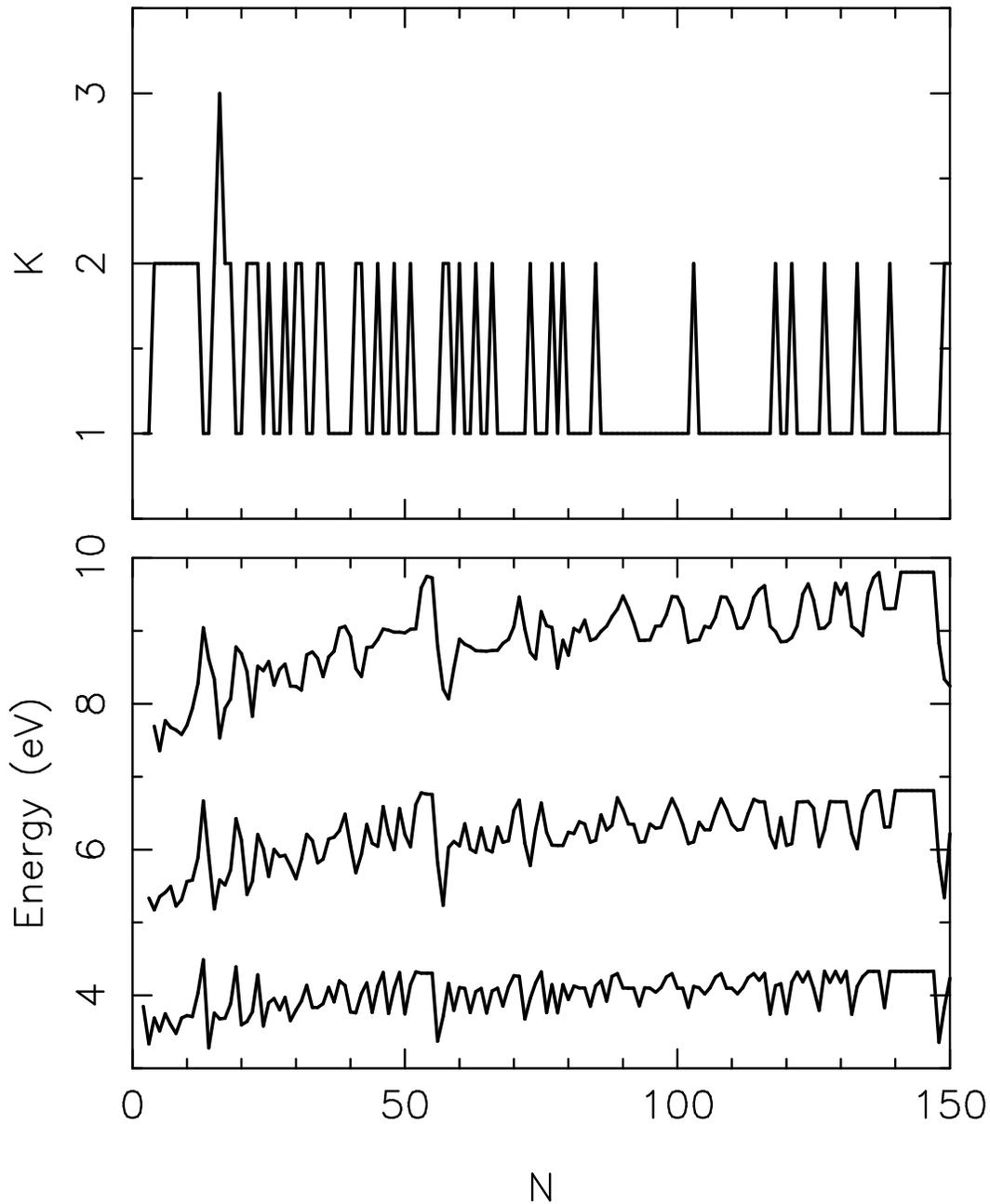}}
\end{picture}
\caption{The fragmentation channels for the Ni$_N$ clusters as functions of $N$.
The upper panel shows the most probable size of the smallest split-off cluster,
whereas the lower panel shows the dissociation energy for splitting off 1 (lowest curve),
2 (middle curve, shifted upwards by 2 eV), or 3 atoms 
(upper curve, shifted upwards by 4 eV).}
\label{fig:fig04}
\end{figure}

Neglecting the $K=0$ solution we found the results shown in Fig.\ \ref{fig:fig04}. Except for
one case, $K$ was found to be either 1 or 2, indicating that the clusters upon fragmentation
will dissociate in a highly asymmetric way by splitting off a single or a pair of atoms.
The exception is that of $N=16$ for which $K=3$ was found as a consequence of the 
particular stability of the $N=13$ cluster. The dissociation energies, also shown in
Fig.\ \ref{fig:fig04}, are seen to lie around 4 eV for $K$ = 1 and 2 and around 5 eV for
$K=3$ for the larger clusters.

Concluding we have presented results of a theoretical study of structural
and energetic properties of Ni$_N$ clusters with all $N$ up to $N=150$. We stress that this study
is the first where a systematic study of energetics, growth, dissociation, and structure 
of not only the absolutely smallest clusters has been
performed without more or less severe assumptions of the structure. Our structure optimization
was performed using our own {\sl Aufbau/Abbau} method, and due to the large number of total-energy
calculations that was necessary we used the approximate EAM for this purpose. 
As the main
result we obtained the structure and the total energy as functions of $N$, and these results were
then subsequently used in identifying particularly stable clusters, in classifying the overall
shape of the clusters, in discussing whether a simple growth mechanism could be applied (this was
found not to be the case), and in discussing possible fragmentation channels. In order to
extract useful information from the obtained structural and energetic
properties we introduced a number of quantitative measures and subsequently
applied them on our results. For the first 
time it was shown that for $73\le N\le 79$ the clusters possess more different types
of structural characteristics, and, moreover, our theoretical study is the first such to show
that for $N=147$ a magic number occurs for a multilayer icosahedron structure. This result is
in perfect agreement with mass-spectroscopical experiments of Pellarin $et$ $al.$\cite{pellarin}

The authors are grateful to Fonds der Chemischen Industrie for very
generous support. This work was supported by the SFB 277 at the 
University of Saarland.

\begin{table}[ht]
\begin{center}
\begin{tabular}{ccccccc}
$N$ & This work & Ref.\ [\onlinecite{calleja}] &
Ref.\ [\onlinecite{michelini}] & Ref.\ [\onlinecite{desmaris}] &
Ref.\ [\onlinecite{krueger}] & Ref.\ [\onlinecite{reuse}] \\ 
\hline
 2 & $D_{\infty h}$ & $D_{\infty h}$ & $D_{\infty h}$ & & & $D_{\infty h}$ \\ 
   &  2.12          &  2.17          &  2.13          & & &   1.99         \\ 
 3 & $D_{3h}$ &  & $D_{3h}$ & & & $C_{2v}$ \\ 
   &  2.25     &  &  2.26     & & & 2.15      \\ 
 4 & $T_d$ &  & $D_{2d}$ ($\sim T_d$) & & & $D_{2d}$ / $D_{4h}$ \\ 
   &  2.32     &  &  2.33     & & & 2.17 / 2.10      \\ 
 5 & $D_{3h}$ &  & $D_{3h}$ & & & $D_{3h}$ \\ 
   &  2.35     &  &  2.36     & & & 2.25      \\ 
 6 & $O_h$ &  & $C_i$ ($\sim O_h$) & & & $D_{4h}$ \\ 
   &  2.36     &  &  2.40     & & & 2.33      \\ 
 7 & $D_{5h}$ &  & & $C_{2v}$ ($\sim D_{5h}$) & &  \\ 
   &  2.39     & &  &  2.28     & &      \\ 
 8 & $D_{2d}$ &  & $T_d$ & $D_2$ ($\sim D_{2d}$) & $D_{2d}$ & $O_h$ \\ 
   &  2.38     & & 2.32 &  2.28     & 2.37 & 2.16     \\ 
13 & $I_h$ & $\sim I_h$ & & & & $\sim I_h$ \\ 
   &  2.36/2.48     & 2.41/2.53 & &  & & 2.23/2.34     \\ 
\end{tabular}
\caption[kurzform]{Our optimized structures in comparison with those of $ab$ $initio$ 
density-functional calculations. In two cases ($N=5$, Ref.\ [\onlinecite{michelini}]
and $N=7$, Ref.\ [\onlinecite{desmaris}])
we compare with the second-lowest structure, 
and in another ($N=4$, Ref.\ [\onlinecite{reuse}]) 
we give their results for the two energetically close minima. The second
row gives averaged bond lengths in \AA. With '$\sim$' we mark approximate symmetries.
The experimental bond length for the
dimer is 2.15$-$2.20 \AA\ (Refs.\ [\onlinecite{exper1,exper2}]).}
\label{comparison}
\end{center}
\end{table}

\unitlength1cm

\unitlength1cm

\unitlength1cm

\unitlength1cm

\unitlength1cm

\end{document}